\documentclass[a4paper,12pt]{article}
\usepackage{jheppub}
\usepackage{graphicx}
\usepackage{epstopdf}

\makeatletter
\def\@fpheader{\relax}
\makeatother

\title{Absence of Black Holes at LHC due to Gravity's Rainbow}

\author[a,b,c]{Ahmed Farag Ali}
\author[d]{Mir Faizal}
\author[e]{Mohammed M. Khalil}

\affiliation[a]{Department of Physics, Florida State University, Tallahassee,\\ FL 32306, USA.}

\affiliation[b]{Center for Fundamental Physics, Zewail City of Science and Technology,\\Giza 12588, Egypt}
\affiliation[c]{Department of Physics, Faculty of Science, Benha University,\\Benha 13518, Egypt}
\affiliation[d]{Department of Physics and Astronomy, University of Waterloo,\\Waterloo, Ontario, N2L 3G1, Canada}
\affiliation[e]{Department of Electrical Engineering, Alexandria University,\\ Alexandria 12544, Egypt}

\emailAdd{ahmed.ali@fsc.bu.edu.eg; afali@fsu.edu}
\emailAdd{f2mir@uwaterloo.ca}
\emailAdd{moh.m.khalil@gmail.com}

\abstract{In this paper, we investigate the effect of Planckian deformation of
	quantum gravity on the production of black holes at colliders using the
	framework of gravity's rainbow. We demonstrate that a black hole remnant exists for
	Schwarzschild black holes in higher dimensions using gravity's rainbow.
	The mass of this remnant is found to be greater than the energy scale at which experiments
	were performed at the LHC.
	We propose this as a possible explanation for the absence of black holes at the LHC.
	Furthermore, we demonstrate that it is possible for black holes in six
	(and higher) dimensions to be produced at energy scales that will be accessible in the near future.}


\arxivnumber{}

\begin{document}

\maketitle
\flushbottom

\section{Introduction}

Black holes are one of the most important objects in quantum gravity. However, there is little hope of detecting a four dimensional
black hole directly in particle accelerators. This is because in order to produce black holes, an energy of the order of the Planck energy ($\sim 10^{19}$ GeV) is needed, and this energy is way beyond what can be achieved in the near future.
However, if large extra dimensions exist,  then there is a hope of observing black holes at colliders, in the near future. This is because the existence of large extra dimensions can lower the effective Planck scale to TeV scales at which experiments can be done \cite{ArkaniHamed:1998rs}. This lowering of Planck scale  occurs in Type I and Type II string theories by localizing the standard model particles on a D-brane, while gravity propagates freely in the higher dimensional bulk. Using this model, it was predicted that due to  this lowering of effective Planck scale, black holes could be produced at the LHC \cite{Banks:1999gd,Giddings:2001bu,Dimopoulos:2001hw,Emparan:2000rs,Meade:2007sz, Antoniadis:1998ig,daRocha:2006ei}. Furthermore, the production  of such black holes would also serve to prove the existence of extra dimensions, and thus provide a strong indication for string theory to be a correct theory describing  the natural world (since string theory is critically based on the existence of higher dimensions).

In the experiments performed at the LHC, no black holes have been detected \cite{Chatrchyan:2012me,Chatrchyan:2012taa}. This result has been interpreted to imply the absence of large extra dimensions, at least at the energy scale at which experiments have been performed at the LHC. However, in this paper, we will demonstrate that these results should rather be interpreted as an indication of a suppression of higher dimensional black hole production due to Planckian deformation of quantum gravity. Since large extra dimensions can lower the effective Planck scale to scales at which such experiments are talking place, it becomes very important to consider the Planckian deformation of quantum gravity. We can implement the Planckian deformation of quantum gravity by introducing rainbow functions in the original classical metric using a formalism called gravity's rainbow.

Gravity's rainbow is motivated by doubly special relativity (DSR), which in turn is motivated by the fact that almost all approaches to quantum gravity suggest that  standard energy-momentum dispersion relation gets deformed near Planck scale. This deformation of the energy-momentum relation has been predicted from
spacetime discreteness \cite{'tHooft:1996uc}, spontaneous symmetry breaking of Lorentz invariance in string field theory \cite{Kostelecky:1988zi}, spacetime foam models \cite{Amelino1997gz}, spin-network in loop quantum gravity (LQG) \cite{Gambini:1998it}, non-commutative geometry \cite{Carroll:2001ws}, and Horava-Lifshitz gravity \cite{Horava:2009uw,Horava:2009if}. As such a deformation of the dispersion relation is a common prediction of various approaches to quantum gravity, we can expect that this will even hold in any quantum theory of gravity. The modification of the dispersion relation generally takes the form,
\begin{equation}
\label{MDR}
E^2f^2(E/E_P)-p^2g^2(E/E_P)=m^2,
\end{equation}
where $E_P$ is the Planck energy, and the functions $f(E/E_P)$ and $g(E/E_P)$ satisfy
\begin{equation}
\lim\limits_{E/E_P\to0} f(E/E_P)=1,\qquad \lim\limits_{E/E_P\to0} g(E/E_P)=1.
\end{equation}

The  modified dispersion relation occurs in DSR because there is a maximum invariant energy scale in addition to the speed of light \cite{AmelinoCamelia:2000mn,Magueijo:2001cr}. The most compelling argument for the existence of such a maximum energy scale comes from string theory. This is because it is not possible to probe spacetime below the string length scale. Thus, string theory comes naturally equipped with a minimum length scale, which can be translated into a maximum  energy scale \cite{Amati:1988tn,Garay:1994en}. DSR can naturally incorporate this  maximum energy scale corresponding to string length scale \cite{Ali:2009zq,Ali:2011fa}. The gravity's rainbow is the generalization of DSR to curved spacetime. This is done by incorporating the functions $ f(E/E_p)$ and $g(E/E_p)$ in general curved spacetime  metric. So, in gravity's rainbow the structure of spacetime depends on the energy used to probe it \cite{Magueijo:2002xx}.

The choice of the rainbow functions $f(E/E_P)$ and $g(E/E_P)$ is important for making predictions.
This choice should be phenomenologically motivated. Different aspects of Gravity's Rainbow with various choices of rainbow functions have been studied in \cite{Galan:2004st,Hackett:2005mb,Garattini:2011hy,Garattini:2013yha,Garattini:2012ec,Garattini:2013psa,Leiva:2008fd,Li:2008gs,Ali:2014cpa,Awad:2013nxa,Barrow:2013gia,Liu:2007fk,Ali:2014aba,Gim:2014ira}. Among these choices, the rainbow functions proposed by Amelino-Camelia, et al. \cite{Amelino1996pj,AmelinoCamelia:1997gz}, are both phenomenologically important and theoretically interesting,
\begin{equation}
\label{rainbowfns}
f\left(E/{E_P}\right)=1,\qquad g\left( E/{E_P} \right)=\sqrt{1-\eta \left(\frac{E}{E_P}\right)^{n}},
\end{equation}
where $n$ is an integer $>0$, and $\eta$ is a constant of order unity, because naturalness says that the parameter is set to be one, unless the observations or measurements prove differently. Besides, in gravity's rainbow, the Planck energy is an invariant scale, and if eta were much greater than one, this would be analogous to reducing the energy scale below the Planck energy.

These rainbow functions lead to the most common form of MDR in the literature. This MDR is compatible with some results from non-critical string theory, loop quantum gravity and $\kappa$-Minkowski non-commutative spacetime \cite{amelino2013}. Furthermore, this MDR was first used to study the possible dispersion of electromagnetic waves from gamma ray bursters \cite{AmelinoCamelia:1997gz}, and it resolved the ultra high energy gamma rays paradox \cite{AmelinoCamelia:2000zs,Kifune:1999ex}. In fact, it was used for  providing an explanation for the 20 TeV gamma rays from the galaxy Markarian 501 \cite{AmelinoCamelia:2000zs,Protheroe:2000hp}. Apart from that, it also provides stringent constraints on deformations of special relativity and Lorentz violations \cite{Aloisio:2000cm,Myers:2003fd}. A detailed  analysis  of the phenomenological aspects of these functions has been  done in \cite{amelino2013}.

An outline of the paper is as follows. In section 2, we review the thermodynamics of higher dimensional Schwarzschild black holes, and in section 3, we study their modified thermodynamics using gravity's rainbow with the rainbow functions Eq. \eqref{rainbowfns}. This is the higher dimensional study of
rainbow Schwarzschild black hole which was studied by one of the authors in \cite{Ali:2014xqa}, and reached the conclusion that black holes end in a remnant. In section 4, we discuss this result and compare it with the energy scale of the LHC. Finally, in section 5, we set bounds on the parameter $\eta$ from LHC experiments. In this paper, we use natural units, in which $c=1$, $\hbar=1$, $G=6.708\times10^{-39}\text{GeV}^{-2}$ and $E_P=1/\sqrt{G}=1.221\times10^{19}\text{GeV}$.

\section{Schwarzschild Black Holes in Higher Dimensions}
In this section, we will review the Schwarzschild black holes in higher dimensions.
This will be used to motivate a similar analysis based on gravity's rainbow, in the next section.
The metric of Schwarzschild black holes in $d$ dimensions takes the form \cite{Emparan:2008eg,Aman:2005xk}
\begin{equation}
\label{metric}
ds^2=-\left(1-\frac{\mu}{r^{d-3}}\right)dt^2+\frac{1}{\left(1-\frac{\mu}{r^{d-3}}\right)}dr^2+r^2d\Omega_{d-2}^2,
\end{equation}
where the mass parameter $\mu$ is given by
\begin{equation}
\mu=\frac{16\pi G_d M}{(d-2)\Omega_{d-2}},
\end{equation}
where $G_d$ is Newton's constant in $d$ dimensions, which is related to the Planck mass $M_P$ via \cite{Dimopoulos:2001hw}
\begin{equation}
G_d=\frac{1}{M_P^{d-2}},
\end{equation}
and $\Omega_{d-2}$ is the volume of the $(d-2)$ unit sphere
\begin{equation}
\Omega_{d-2}=\frac{2\pi^{\frac{d-1}{2}}}{\Gamma\left(\frac{d-1}{2}\right)}.
\end{equation}
The horizon radius $r_h$ is evaluated by solving $(1-\mu/r_h^{d-3})=0$ leading to
\begin{equation}
\label{radius}
r_h=\mu^{\frac{1}{d-3}}=\frac{1
}{\sqrt{\pi}}\left(\frac{8M\Gamma\left(\frac{d-1}{2}\right)}{M_P^{d-2}(d-2)}\right)^{\frac{1}{d-3}}.
\end{equation}

The Hawking temperature can be calculated via the relation \cite{Angheben:2005rm}
\begin{equation}
\label{temp}
T=\frac{1}{4\pi}\sqrt{A_{,r}(r_h)B_{,r}(r_h)}.
\end{equation}
This relation applies to any spherically symmetric black hole with a metric of the form
\begin{equation}
ds^2=-A(r)dt^2+\frac{1}{B(r)}dr^2+h_{ij}dx^idx^j.
\end{equation}
From the Schwarzschild metric in Eq. \eqref{metric}, $A(r)=B(r)=1-\mu/r^{d-3}$. Thus, we get the temperature
\begin{equation}
T=\frac{d-3}{4\pi r_h},
\end{equation}
and when we substitute the value of $r_h$ from Eq. \eqref{radius} we get \cite{Cavaglia:2003qk}
\begin{equation}
\label{temperature}
T=\frac{d-3}{4\sqrt{\pi}}\left(\frac{M_P^{d-2}(d-2)}{8M \Gamma\left(\frac{d-1}{2}\right)}\right)^{\frac{1}{d-3}}.
\end{equation}
Since $d\geq4$, the temperature goes to infinity as $M\to0$. Figure \ref{fig:temp} is a plot of this equation for $d=4, d=6,$ and $d=10$,  with the generic values $n=4$, $\eta=1$, and $M_P=1$; different values lead to the same qualitative behavior.

The black hole entropy can be calculated from the first law of black hole thermodynamics $dM=TdS$ leading to
\begin{equation}
\label{entropy}
S=\int \frac{1}{T}dM =\frac{4\sqrt{\pi}}{d-2}\left(\frac{8\Gamma\left(\frac{d-1}{2}\right)}{d-2}\right)^{\frac{1}{d-3}}\left(\frac{M}{M_P}\right)^{\frac{d-
2}{d-3}},
\end{equation}
which goes to zero as $M\to0$.

The specific heat capacity is calculated from the relation
\begin{equation}
\label{heatcap}
C=T\frac{\partial S}{\partial T}=\frac{\partial M}{\partial T}.
\end{equation}
By differentiating the temperature from Eq. \eqref{temperature} with respect to $M$ we get
\begin{equation}
\label{capacity}
C=-4\sqrt{\pi}\left(\frac{8\Gamma\left(\frac{d-1}{2}\right)}{d-2}\right)^{\frac{1}{d-3}}\left(\frac{M}{M_P}\right)^{\frac{d-2}{d-3}}.
\end{equation}

The emission rate (the energy radiated per unit time) can be calculated from the temperature using the Stefan-Boltzmann law assuming the energy loss is dominated by photons. In $m$-dimensional brane the emission rate of a black body with temperature $T$ and surface area $A_m$ is given by \cite{Emparan:2000rs}
\begin{equation}
\frac{dM}{dt}=\sigma_m A_m T^m,
\end{equation}
where $\sigma_m$ is the Stefan-Boltzmann constant in $m$ dimensions. Since black holes are radiating mainly
on the brane \cite{Emparan:2000rs}, so using $m=4$ as in \cite{Cavaglia:2003qk}, and since $A\propto M^{\frac{2}{d-3}}$ and from Eq. \eqref{temperature} $T \propto M^{\frac{-1}{d-3}}$  we get that
\begin{equation}
\label{rate}
\frac{dM}{dt}\propto M^{\frac{-2}{d-3}}.
\end{equation}
The exact form can be found in \cite{Emparan:2000rs,Cavaglia:2003qk}.

From the relations Eq. \eqref{temperature}, \eqref{entropy}, \eqref{capacity}, and \eqref{rate}, we see that when the black hole evaporates  and its mass goes to zero, the temperature and emission rate go to infinity, while the entropy and heat capacity vanish. This means that the black hole reaches a stage of \emph{catastrophic evaporation} as the black hole mass approaches zero, and this definitely needs a resolution. This problem has been tackled in \cite{Adler:2001vs}, and it has been resolved
by considering the generalized uncertainty principle \cite{Amati:1988tn} instead of the standard uncertainty principle, and in this picture, black holes end at a remnant that does not exchange hawking radiation with the surroundings. Similar conclusion was obtained by one of the authors in \cite{Ali:2014xqa}, in which it was studied the thermodynamics of Schwarzschild black holes in the context of gravity's rainbow, and it was found that the rainbow black hole ends at a remnant at which the specific heat vanishes and hence the catastrophic behavior is again resolved but this time in the context of gravity's rainbow. In the next section, we shall extend this study into extra dimensions to investigate the phenomenological implications
on the productions of black holes at TeV scales.

\section{Schwarzschild Black Holes in Gravity's Rainbow}
In this section, we will analyze the Schwarzschild black hole in higher dimensions using gravity's rainbow. The four dimensional Schwarzschild black hole has been analyzed in gravity's rainbow \cite{Ali:2014xqa}, and it was found that a remnant forms. In this section, we extend this analysis into higher dimensional Schwarzschild black holes.
In gravity's rainbow, the geometry of spacetime depends on the energy $E$ of the particle used to probe it, and  so,  the rainbow modified metric can be written as   \cite{Magueijo:2002xx}
\begin{equation}
\label{rainmetric}
g(E)=\eta^{ab}e_a(E)\otimes e_b(E).
\end{equation}
The energy dependence of the frame fields can be written as
\begin{equation}
e_0(E)=\frac{1}{f(E/E_P)}\tilde{e}_0, \qquad
e_i(E)=\frac{1}{g(E/E_P)}\tilde{e}_i,
\end{equation}
where the tilde quantities refer to the energy independent frame fields. So,  we can write the  modified Schwarzschild metric as \cite{Magueijo:2002xx,Liu:2014ema}
\begin{equation}
ds^2=-\frac{A(r)}{f(E)^2}dt^2+\frac{1}{g(E)^2B(r)}dr^2+\frac{r^2}{g(E)^2}d\Omega_{d-2}^2.
\end{equation}
where $f(E)$ and $g(E)$ are the rainbow functions used in the MDR given in Eq. \eqref{MDR}.

Thus, the modified temperature can be calculated from Eq. \eqref{temp} with the change $A(r)\to A(r)/f(E)^2$ and $B(r)\to B(r)g(E)^2$ leading to
\begin{equation}
T'=T\frac{g(E)}{f(E)}=T\sqrt{1-\eta \left(\frac{E}{E_P}\right)^{n}},
\end{equation}
where we used the rainbow functions from Eq. \eqref{rainbowfns}.
According to \cite{Adler:2001vs, Cavaglia:2003qk, Medved:2004yu, AmelinoCamelia:2004xx}, the uncertainty principle $\Delta p\geq 1/\Delta x$ can be translated to a lower bound on the energy $E\geq 1/\Delta x$ of a particle emitted in Hawking radiation, and the value of the uncertainty in position can be taken to be the event horizon radius. Hence,
\begin{equation}
E\geq \frac{1}{\Delta x} \approx \frac{1}{r_h}.
\end{equation}
The temperature becomes
\begin{align}
\label{modtemp}
T'&=\frac{d-3}{4\pi r_h} \sqrt{1-\eta \left(\frac{1}{r_h M_P}\right)^n} \nonumber\\
&=\frac{d-3}{4\sqrt{\pi}}\left(\frac{M_P^{d-2}(d-2)}{8M \Gamma \left(\frac{d-1}{2}\right)} \right)^{\frac{1}{d-3}} \sqrt{1-\eta \pi^{\frac{n}{2}}\left(\frac{M_P (d-2)}{8M\Gamma\left(\frac{d-1}{2}\right)}\right)^{\frac{n}{d-3}}},
\end{align}
where we used $E_P=M_P$ in natural units.

\begin{figure}[th]
	\centering
	\begin{minipage}[b]{0.45\linewidth}
		\includegraphics[width=\linewidth]{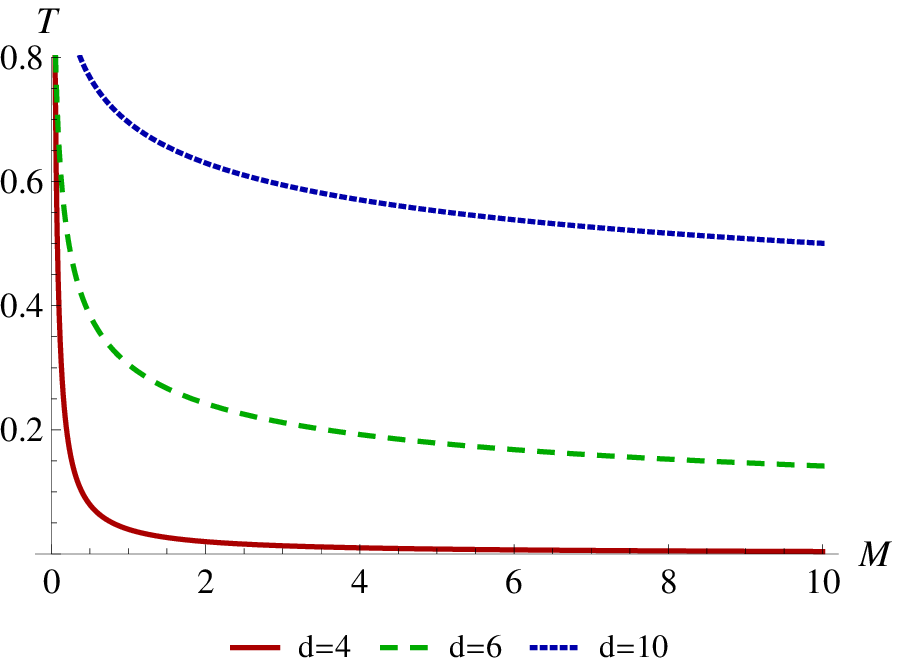}
		\caption{\label{fig:temp} Standard temperature of Schwarzschild black hole for $d=4, d=6$ and $d=10$.}
	\end{minipage}
	\quad
	\begin{minipage}[b]{0.45\linewidth}
		\includegraphics[width=\linewidth]{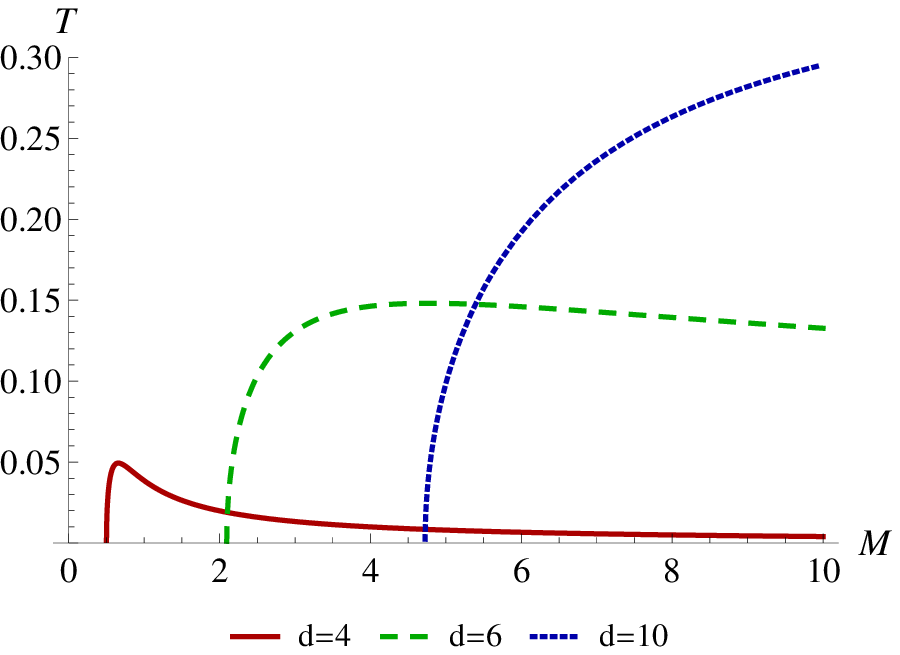}
		\caption{\label{fig:modtemp} Modified temperature due to gravity's rainbow for $d=4, d=6$ and $d=10$.}
	\end{minipage}
\end{figure}

From Eq. \eqref{modtemp}, it is clear that the temperature goes to zero at $r_h=\eta^{\frac{1}{n}}/M_P$, and below this value the temperature has no physical meaning. This minimum horizon radius corresponds to the minimum mass
\begin{equation}
\label{Mmin}
M_{min}=\frac{d-2}{8\Gamma\left(\frac{d-1}{2}\right)}\pi^{\frac{d-3}{2}}\eta^{\frac{d-3}{n}} M_P.
\end{equation}
This implies that the black hole ends in a \emph{remnant}. Figure \ref{fig:modtemp} is a plot of Eq. \eqref{modtemp} for $d=4, d=6,$ and $d=10$.

The entropy can be calculated from the first law of black hole thermodynamics using the modified temperature from Eq.  \eqref{modtemp}
\begin{align}
S'=\int\frac{1}{T'}dM=&\frac{4\sqrt{\pi}}{d-3}\left(\frac{8\Gamma\left(\frac{d-1}{2}\right)}{M_P^{d-2}(d-2)}\right)^{\frac{1}{d-3}} \nonumber\\ &\int\frac{M^{\frac{1}{d-3}}}{\sqrt{1-\eta\pi^{\frac{n}{2}} \left(\frac{M_P(d-2)}{8M\Gamma\left(\frac{d-1}{2}\right)}\right)^{\frac{n}{d-3}}}}dM
\end{align}
This integral cannot be evaluated exactly for general $n$ and $d$, but taking as an example $d=4$ and $n=4$ we get
\begin{equation}
S'=\frac{4\pi M^2}{M_P^2}\sqrt{1-\eta\left(\frac{M_P}{2M}\right)^4},
\end{equation}
which is the same as the expression derived in \cite{Ali:2014xqa}. Taking as another example $d=5$ and $n=2$ we get
\begin{equation}
S'=\frac{1}{3}\sqrt{\frac{\pi M}{3M_P^3}}(4M+3\pi\eta M_P)\sqrt{8-\frac{3\pi\eta M_P}{M}}.
\end{equation}

\begin{figure}[t]
	\centering
	\begin{minipage}[b]{0.45\linewidth}
		\includegraphics[width=\linewidth]{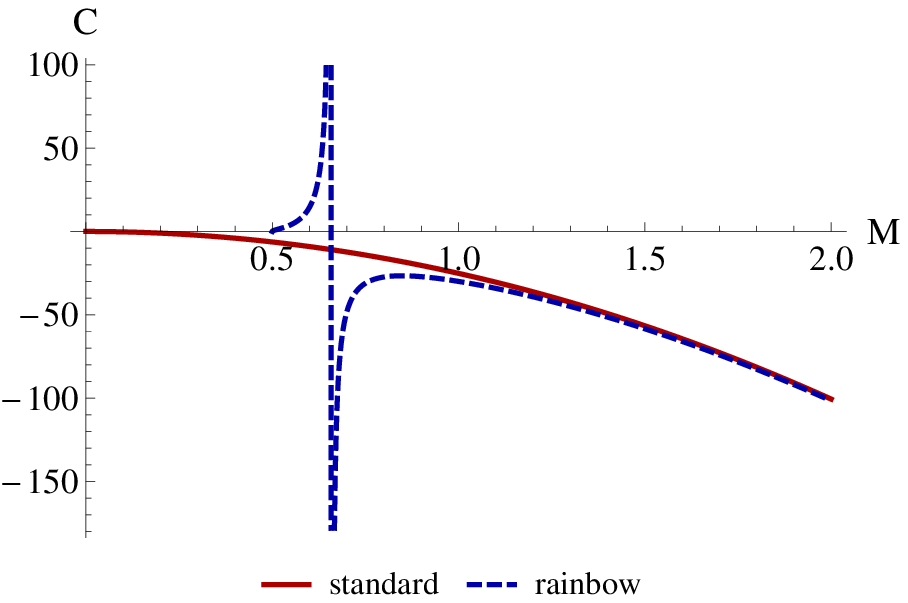}
		\caption{\label{fig:cap4} Standard and modified specific heat capacity of Schwarzschild black hole for $d=4$.}
	\end{minipage}
	\quad
	\begin{minipage}[b]{0.45\linewidth}
		\includegraphics[width=\linewidth]{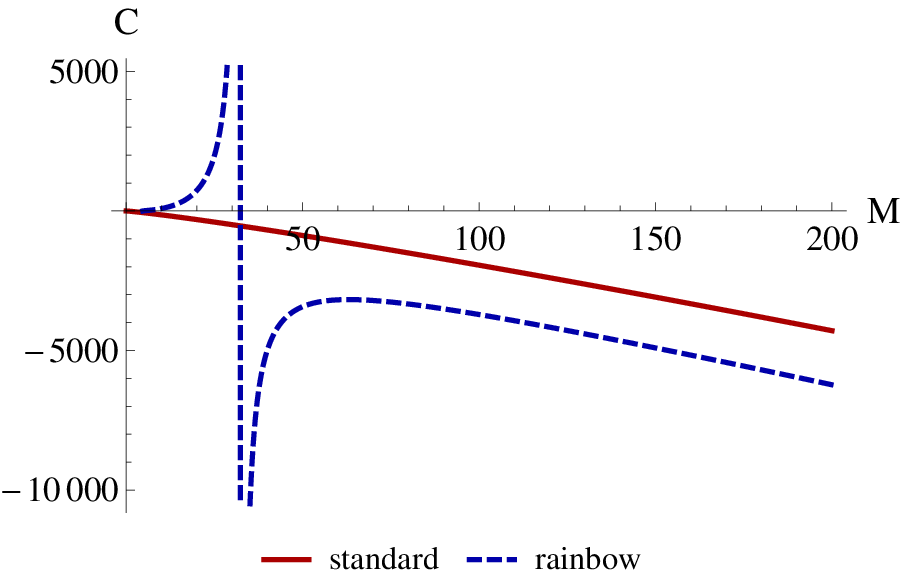}
		\caption{\label{fig:cap10} Standard and modified specific heat capacity of Schwarzschild black hole for $d=10$.}
	\end{minipage}
\end{figure}

The heat capacity can be calculated from Eq. \eqref{heatcap} with the modified temperature in Eq. \eqref{modtemp}, and we get
\begin{equation}
C'=-4\sqrt{\pi}\left(\frac{8M^{d-2}\Gamma\left(\frac{d-1}{2}\right)}{M_P^{d-2}(d-2)}\right)^{\frac{1}{d-3}}
\frac{\sqrt{1-\eta\pi^{\frac{n}{2}}\left(\frac{M_P(d-2)}{8M\Gamma\left(\frac{d-1}{2}\right)}\right)^{\frac{n}{d-3}}}}
{1-\frac{n+2}{2}\eta\pi^{\frac{n}{2}} \left(\frac{M_P(d-2)}{8M\Gamma\left(\frac{d-1}{2}\right)}\right)^{\frac{n}{d-3}}}.
\end{equation}
Figures \ref{fig:cap4} and \ref{fig:cap10} are plots of the heat capacity for $d=4$ and $d=10$ respectively. We see that the modified heat capacity diverges at a value where the temperature is maximum, then goes to zero at the minimum mass given by Eq. \eqref{Mmin}. The zero value of the heat capacity means the black hole cannot exchange heat with the surrounding space, and hence predicting the existence of a remnant.

The emission rate is proportional to $T^4$, which means that from the modified temperature in Eq. \eqref{modtemp}, the modified emission rate is
\begin{equation}
\label{emission}
\left(\frac{dM}{dt}\right)_{rainbow}=\frac{dM}{dt} \left(1-\eta\left(\frac{1}{r_h M_P}\right)^n\right)^2,
\end{equation}
which also goes to zero at $r_h=\eta^{\frac{1}{n}}/M_P$.

From the calculations in this section, we conclude that in gravity's rainbow black holes reach a remnant near the Planck scale. In the next section, we investigate whether black hole remnants can be detected in the LHC.

\section{Black Hole Production at Colliders}
In the last section, we found that in gravity's rainbow, black holes end up in a remnant with the mass in Eq. \eqref{Mmin}, which we reproduce here for convenience,
\begin{equation}
M_{min}=\frac{d-2}{8\Gamma\left(\frac{d-1}{2}\right)}\pi^{\frac{d-3}{2}}\eta^{\frac{d-3}{n}} M_P.
\end{equation}
From this minimum mass, we can calculate the minimum energy needed to form black holes in a collider, such as the LHC.
In the ADD model \cite{ArkaniHamed:1998rs}, the reduced Planck constant $M_P$ in extra dimensions is related to the 4D Planck mass $M_{P(4)}\sim 10^{19}$ GeV via
\begin{equation}
\label{Mp}
M_{P(4)}^2=R^{d-4} M_P^{d-2}.
\end{equation}
where $R$ is the size of the compactified extra dimensions. Fixing $M_P$ at around the electroweak scale $\sim$TeV, and using Eq.  \eqref{Mp}, we obtain
$
d=5,\,  6,...,\,  10 \quad \to \quad R\sim 10^9\text{km}, \,  0.5\text{mm},..., \,  0.1 \text{MeV}^{-1}$  \cite{Beringer:1900zz}.
Thus, $d=5$ is clearly ruled out, but not $d\geq6$.

When we use the latest experimental limits on $M_P$ from Ref. \cite{Chatrchyan:2012me}, and assume that the rainbow parameter $\eta=1$, we obtain  the results given in Table 1. We see that in $d=6$, black holes can form only at energies not less than $9.5$ TeV, and in $d=10$ the minimal mass is $11.9$ TeV. This energy scale is larger than the energy scale of the current runs of the LHC, which explains why they were not detected in the LHC. Previous work based on theories with large extra dimensions predicted the possibility of forming black holes at energy scales of a few TeVs \cite{Giddings:2001bu,Dimopoulos:2001hw,Emparan:2000rs,Cavaglia:2003qk}, which has not been experimentally observed at the Compact Muon Solenoid (CMS) detector in LHC where experiments are excluding semiclassical and quantum black holes with masses below $3.8$ to $5.3$ TeV \cite{Chatrchyan:2012me,Chatrchyan:2012taa}. 
We also note that our results may ameliorate the ranges of masses of black holes that has been predicted in the earlier work in Fig. (2) in \cite{Dimopoulos:2001hw} that gave a wide range between around 1.5 TeV and 10 TeV. 

By considering our proposed approach of studying black holes in the context of gravity's rainbow, we may justify why higher energy scales are needed to form black holes. Furthermore, this energy scale will be accessible in the near future. If black holes were produced in future colliders, it will need a collision center-of-mass energy greater than the minimal mass. The emitted radiation from the evaporation will be smaller than the standard case (Eq. \eqref{emission}), and the emission will stop when the black hole reaches the remnant mass. This will lead to the detection of a missing energy of the order of the remnant mass.

The total cross section of a collision that produces a black hole can be estimated by \cite{Dimopoulos:2001hw}
\begin{equation}
\sigma(M)\approx\pi r_h^2=\left(\frac{8M\Gamma\left(\frac{d-1}{2}\right)}{M_P^{d-2}(d-2)}\right)^{\frac{2}{d-3}},
\end{equation}
and the differential cross section
\begin{equation}
\frac{d\sigma}{dM}=\frac{2}{(d-3)M}\left(\frac{8M\Gamma\left(\frac{d-1}{2}\right)}{M_P^{d-2}(d-2)}\right)^{\frac{2}{d-3}}.
\end{equation}
The maximum number of expected events per second is given by
\begin{equation}
\frac{dR}{dt}=L\sigma.
\end{equation}
For the LHC, the luminosity  $L\approx 10^{34} \text{cm}^{-2}\text{s}^{-1}$, and the total center of mass energy is currently 7 TeV, but can be increased up to 14 TeV in future runs.

\begin{table}[h]
\centering
\begin{tabular}{cccccc}
	\hline
	$d$	& $M_P$ [TeV] & $M_{min}$ [TeV] & $\sigma$ [pb] & $\frac{d\sigma}{dM}$ [pb/100 GeV] & $\frac{dR}{dt}$ [events/s]\\ \hline
	6	& 4.54 & 9.5 & 59.4 & 0.42 & 0.59 \\ 
	7	& 3.51 & 10.8 & 99.4 & 0.46 & 0.99 \\ 
	8	& 2.98 & 11.8 & 137.8 & 0.47 & 1.38 \\ 
	9	& 2.71 & 12.3 & 166.7 & 0.45 & 1.67 \\ 
	10	& 2.51 & 11.9 & 194.3 & 0.47 & 1.94 \\ \hline
\end{tabular} 
\caption{Mass of the black hole remnant, cross section, differential cross section, and the maximum number of expected events per second in different dimensions. The values of $M_P$ are from \cite{Chatrchyan:2012me}.}
\label{table1}
\end{table}

Table \ref{table1} includes the estimated cross section, differential cross section, and the maximum number of expected events per second. For comparison, the cross section of the Higgs boson is approximately $50$ fb, and the number of events per second is $5\times 10^{-4}$. This means that for a collision with energies higher than the remnant mass of the black holes, the production of black holes could be more than that of the Higgs. 

However, the values of the cross section in Table \ref{table1} will decease if one takes into account that only a fraction of the energy in a $pp$ collision is achieved in a parton-parton scattering \cite{Dimopoulos:2001hw}. In addition, the minimal mass is sensitive to the value of the parameter $\eta$. For example, for $\eta=1.1$ and $d=6$, $M_{min}=10.97$ TeV. Also, for $\eta=2$ and $d=6$,  $M_{min}=26.9$ TeV. 
Thus, to determine the expected number of produced black holes accurately, we need better constraints on the parameter $\eta$ from other experiments  \cite{Ali:2014aba}, and simulate the production and decay of black hole remnants as was done in \cite{Bellagamba:2012wz, Alberghi:2013hca}.

\section{Bounds on $\eta$}
In the previous section, we used the value $\eta=1$ to calculate the expected mass of the remnant. We could do the reverse and constrain the value of the parameter $\eta$ from the measurements of no black holes at LHC up to 5.3 TeV \cite{Aad:2013gma}. From Eq. \eqref{Mmin}, $M_{min}>5.3TeV$,
\begin{equation}
5.3\text{TeV}>\frac{d-2}{8\Gamma\left(\frac{d-1}{2}\right)}\pi^{\frac{d-3}{2}}\eta^{\frac{d-3}{n}} M_P,
\end{equation}
which constrains $\eta$ by
\begin{equation}
\eta > \left(\frac{5.3\times 8\Gamma\left(\frac{d-1}{2}\right)} {(d-2)\pi^{\frac{d-3}{2}}M_P}\right)^{\frac{n}{d-3}}.
\end{equation}

\begin{table}[b]
	\centering
	\begin{tabular}{c|ccccc}
		\hline\rule{0pt}{3ex} $d$ & 6 & 7 & 8 & 9 & 10 \\
		\hline\rule{0pt}{3ex} $\eta > $ & 0.68 & 0.70 & 0.73 & 0.76 & 0.79  \\
		\hline
	\end{tabular}
	\caption{Mass of the black hole remnant vs the parameter $\eta$ for different dimensions.}
	\label{table2}
\end{table}

\begin{figure}[t]
	\centering
	\includegraphics[width=0.45\linewidth]{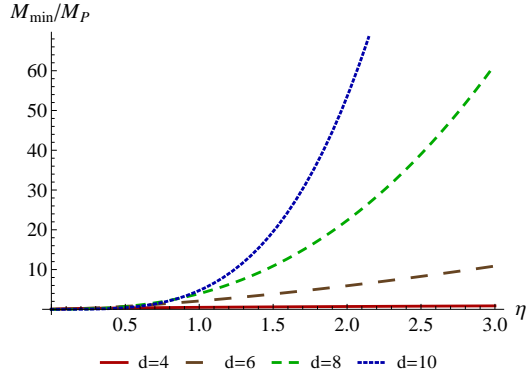}
	\caption{\label{fig:eta} Minimal mass vs the parameter $\eta$ for $d=4,6,8,10$.}
\end{figure}

Table \ref{table2} shows the bounds on $\eta$ in different dimensions, and fig \ref{fig:eta} is a plot for the minimal mass vs $\eta$. To our knowledge, the best upper bound on $\eta$ in the context of gravity's rainbow is $10^5$, but can be reduced by 4 orders in the next few years from tests of the weak equivalence principle \cite{Ali:2014aba}.  Combining these two bounds supports the assumption that $\eta \sim 1$.

\section{Conclusions}
In this paper, we have analyzed higher dimensional Schwarzschild black holes in gravity's rainbow. It was expected that black holes will be detected at LHC if large extra dimensions existed. This was because the existence of extra dimensions would lower the effective Planck mass to TeV scale (i.e LHC energy scale). The absence of any black hole at LHC could thus be interpreted as the absence of large extra dimensions, at least at the energy scale of the LHC. However, we argued that black holes were not detected due to Planckian deformation of quantum gravity, which was not taken into account. As the effective Planck scale was reduced due to the existence of large extra dimensions, it is important that these effects are taken into account. When we did that using gravity's rainbow, we found that the energy needed to form black holes is larger than the energy scale of the LHC, but is within reach of the next particle colliders.

It may be noted that such a suppression was predicted in the framework of generalized uncertainty principle in
\cite{Cavaglia:2003qk,Ali:2012mt,Hossenfelder:2004ze}.
The fact that the generalized uncertainty principle can lead to a deformed dispersion relation suggests that this might be a general feature of theories with modified dispersion relation. It would be interesting to analyze this relation in more details. It is worth mentioning, suppression on black hole masses at Tetra scale was studied in non-commutative geometry  \cite{Nicolini:2011nz,Mureika:2011hg}. Useful reviews on the remnant of black holes in the framework of noncommutative geometry can be found in \cite{Nicolini:2008aj,Bleicher:2014laa}.

Apart from this phenomenological result, it was demonstrated that a black hole remnant will form for higher dimensional Schwarzschild black holes. Such a remnant forms for a four dimensional Schwarzschild black hole \cite{Ali:2014xqa}. In fact, recently it was demonstrated that a remnant also forms for black rings \cite{Ali:2014yea}. These are strong indications that a remnant might form for all black objects, in gravity's rainbow. It will be appropriate to
extend the investigation into dark matter, cosmological constant, etc in the context of gravity's rainbow. We hope to report on these in the future.

\subsection*{Acknowledgments}
The research of AFA is supported by Benha University (www.bu.edu.eg)
and CFP in Zewail City.

\bibliography{BH-LHC}
\bibliographystyle{JHEP}

\end{document}